%% file: ms.tex
\documentclass[10pt,journal,letterpaper]{IEEEtran}
\IEEEoverridecommandlockouts

\usepackage[english]{babel}
\usepackage[utf8]{inputenc}
\usepackage[T1]{fontenc}

\usepackage{amssymb}
\usepackage{amsmath}
\usepackage{graphicx}
\graphicspath{{imgs/}{figures/}{plots/}}
\usepackage[scaled]{beramono}
\usepackage[numbers]{natbib}
\usepackage{paralist}
\usepackage{textcomp}

\usepackage[inline]{enumitem}
\usepackage{algorithm}
\usepackage[noend]{algpseudocode}

\usepackage{array}
\usepackage{subfig}
\usepackage{booktabs}
\usepackage{datetime}
\usepackage{caption}
\usepackage{etoolbox}
\usepackage[detect-all]{siunitx}
\usepackage{blkarray}

\usepackage{acro}
\DeclareAcronym{NC}{short=NC, long=Network Calculus}
\DeclareAcronym{DNC}{short=DNC, long=Deterministic Network Calculus}
\DeclareAcronym{TFA}{short=TFA, long=Total Flow Analysis}
\DeclareAcronym{SFA}{short=SFA, long=Separate Flow Analysis}
\DeclareAcronym{PMOO}{short=PMOO, long=Pay Multiplexing Only Once}
\DeclareAcronym{TSN}{short=TSN, long=Time Sensitive Networking}
\DeclareAcronym{DiffNC}{short=DiffNC, long=Differential Network Calculus}
\DeclareAcronym{AD}{short=AD, long=automatic differentiation}
\DeclareAcronym{SQP}{short=SQP, long=sequential quadratic programming}
\DeclareAcronym{SLSQP}{short=SLSQP, long=sequential least squares quadratic programming}
\DeclareAcronym{IPOPT}{short=IPOPT, long=Interior Point OPTimizer}
\DeclareAcronym{MMA}{short=MMA, long=method of moving asymptotes}
\DeclareAcronym{CCSA}{short=CCSA, long=conservative convex separable approximation}
\DeclareAcronym{BONMIN}{short=BONMIN, long=Basic Open-source Nonlinear Mixed INteger programming}
\DeclareAcronym{GNN}{short=GNN, long=graph neural network}
\DeclareAcronym{NN}{short=NN, long=neural network}
\DeclareAcronym{LP}{short=LP, long=linear program}
\DeclareAcronym{NLP}{short=NLP, long=nonlinear programming}
\DeclareAcronym{ILP}{short=ILP, long=integer linear programming}
\DeclareAcronym{MINLP}{short=MINLP, long=mixed-integer nonlinear programming}
\DeclareAcronym{ML}{short=ML, long=machine learning}
\DeclareAcronym{CAS}{short=CAS, long=computer algebra system}
\DeclareAcronym{TDMA}{short=TDMA, long=time-division multiple access}
\DeclareAcronym{FIFO}{short=FIFO, long=first-in first-out}
\DeclareAcronym{ULP}{short=ULP, long=unique linear program}
\DeclareAcronym{AFDX}{short=AFDX, long=Avionics Full-Duplex Switched Ethernet}
\DeclareAcronym{VM}{short=VM, long=virtual machine}

\usepackage{hyphenat}
\newcommand\R{\mathbb{R}} %
\newcommand{\TP}[1]{10\textsuperscript{#1}}

\usepackage{mathtools}

\newtheorem{theorem}{Theorem}
\newtheorem{definition}{Definition}

\newtheorem{lemma}{Lemma}

\author{Fabien Geyer and Steffen Bondorf%
\IEEEcompsocitemizethanks{%
	\IEEEcompsocthanksitem The preliminary version of this article is published in \cite{GeyerBondorf_INFOCOM2022} [DOI: 10.1109/INFOCOM48880.2022.9796777]. \textit{(Corresponding author: Fabien Geyer.)}
	\IEEEcompsocthanksitem F. Geyer is with Airbus Central Research \& Technology, Munich, Germany (email: fabien.geyer@airbus.com).
	\IEEEcompsocthanksitem S. Bondorf is with the Faculty of Computer Science, Ruhr University Bochum, Germany (email: steffen.bondorf@rub.de).
}}

\usepackage{nameref}
\usepackage[capitalize,noabbrev]{cleveref}

\begin{document}

\title{Differentiable Programming \& \acl{NC}:\\
Configuration Synthesis under Delay Constraints}

\maketitle
\acresetall

\input{abstract}

\IEEEdisplaynontitleabstractindextext
\IEEEpeerreviewmaketitle

\acresetall
\input{introduction}

\input{related_work}

\input{netcal_background}

\input{diffnc}

\input{linear_formulation}

\input{evaluation}

\acresetall
\input{conclusion}

\footnotesize
\yyyymmdddate
\bibliographystyle{IEEEtran}
\bibliography{IEEEabrv,biblio}

\input{bibliographies}

\end{document}

%% file: abstract.tex
\IEEEtitleabstractindextext{%
\begin{abstract}
With the advent of standards for deterministic network behavior, %
synthesizing network designs under delay constraints becomes the natural next task to tackle. %
\ac{NC} has become a key method for validating industrial networks, as it computes formally verified end-to-end delay bounds. %
However, analyses from the \acs{NC} framework have been designed to bound the delay of one flow at a time.
Attempts to use classical analyses to derive a network configuration have shown that this approach is poorly suited to practical use cases. %
Consider finding a delay-optimal routing configuration: %
one model had to be created for each routing alternative, then each flow delay had to be bounded, and then the bounds had to be compared to the given constraints.
To overcome this three-step process, we introduce \acl{DiffNC}.
We extend \acs{NC} to allow the differentiation of delay bounds w.r.t. to a wide range of network parameters -- such as flow paths or priority.
This opens up \acs{NC} to a class of efficient nonlinear optimization techniques that exploit the gradient of the delay bound.
Our numerical evaluation on the routing and priority assignment problem shows that our novel method can synthesize flow paths and priorities in a matter of seconds, outperforming existing methods by several orders of magnitude.
\end{abstract}

}

%% file: introduction.tex
\section{Introduction}

\IEEEPARstart{W}{ith} the recent development of networking solutions with stringent reliability and safety requirements (such as IEEE \ac{TSN}), the formal verification and optimization of safety-critical networks has become an important step in the design process in various industries \cite{Geyer2016}.
While the use of mathematical models and the formalization of end-to-end delay bounds has now become common practice, the optimization and fine-tuning of networks under such formulations remains a challenging task.
The main difficulty arises from the inherent combinatorial property of the formal models which are oftentimes nonlinear as well.
This makes them hard problems to solve in polynomial time.
Previous attempts have often been limited to small networks.

In this article, we propose a novel approach for modeling, optimizing, and synthesizing networks under hard end-to-end delay constraints, able to scale to networks of realistic sizes.
We present an approach able to efficiently synthesize parameters; illustrated on flow paths and flow priorities, yet extendable to. e.g., schedulers' parameters, too.
We bound end-to-end delays using \ac{NC} based on the (min,plus) algebra~\cite{LeBoudec2001}.
While this method is commonly used in some industries to formally validate delay requirements, 
it is rarely used for synthesis or as a design tool.
Existing \ac{NC} analyses have been designed to analyze already completed network designs, making them only suitable for design space exploration that enumerates and ranks different designs.

We present an extension of \ac{NC} called \ac{DiffNC}.
We formally show that under the assumptions traditionally used for validating industrial networks (i.e., token-bucket and rate-latency curves), the delay bound of a flow computed using the (min,plus) algebra is differentiable according to the parameters of the different curves in the network.
This enables a wide range of applications, including gradient-based \ac{NLP}.
Via variable relaxation, we demonstrate that traditional \ac{NLP} methods based on Newton's method can efficiently solve the aforementioned network optimization problems -- and synthesize configurations.
We show that these optimization methods are highly efficient, scale well and provide the best solutions, making them applicable to networks of sizes found in the industry.

In the realm of \ac{NC}, previous works already formalized \ac{NC} as an optimization problem, by proposing a formulation of the end-to-end delay bounds as an \ac{LP}~\cite{Bouillard2010}.
This approach is able to achieve tight delay bounds.
We illustrate that this \ac{LP} formulation can also be extended to  %
multiple flows in its objective function. 
It can be used to optimize paths of flows, but the objective function suffers from poor expressiveness for some important types of constraints. 
In addition, we show that it suffers from poor scalability, requiring more than an hour of computation even on relatively small networks.
These limitations make the approach unsuitable for realistic problems.

Our proposed approach has the following advantages.
First, we use an existing \ac{NC} analysis to derive a (min,plus)-algebraic term bounding the delay, yet with integer variables encoding alternatives such as potential flow paths or flow priority. 
Then, for finding the best alternative with \ac{NLP}, we can include nonlinear constraints and nonlinear objective functions, allowing for concepts like utility functions \cite{Kelly1998} on the delay bounds.

To solve the \ac{NLP}, we chose the Frank-Wolfe algorithm, an efficient \ac{NLP} based on the well-known gradient descent algorithm.
We show that it can be used to find a good solution in terms of optimality in a short amount of time.
In our numerical evaluation, we demonstrate that we are able to optimize the \ac{AFDX} network used in the Airbus A350 in a matter of seconds, with better optimality than previous works.
We thus illustrate that our approach is scalable to real networks with more than \num{1000} flows.
Our implementation is based on efficient \ac{CAS} and \ac{AD}, which allows us to efficiently compute the end-to-end delay bounds and their gradient without paying dearly in terms of computation time.

This article is organized as follows:
\cref{sec:related_work} presents the related work, followed by \ac{NC} in \cref{sec:dncintro}.
\Cref{sec:diffnc} presents the mathematical foundations of \ac{DiffNC} and suitable network optimization problems,
followed by our \ac{DiffNC} implementation in \cref{sec:diffnc_ad}.
\cref{sec:optlp} extends existing \ac{LP}-based \ac{NC} to compete with \ac{DiffNC}.
We numerically evaluate and compare \ac{DiffNC} against other optimization methods in \cref{sec:numerical_evaluation}.
Finally, \cref{sec:conclusion} concludes the article.

%% file: related_work.tex
\section{Related work}
\label{sec:related_work}

\subsubsection*{Optimization with delay bounds}

Various works already investigated delay bound minimization.
\cite{Bouillard2008b} proposed one of the early works on route optimization based on \ac{NC} by using shortest path on graphs with weights set according to delay bounds.
While this approach was shown to be efficient, it is limited to the optimization of a single flow's route, restricting its use when multiple routes need to be optimized.

In \cite{Cattelan2017}, various iterative algorithms for rerouting flows to minimize tail delays were detailed.
These \ac{NC}-based algorithms appeared to scale to realistic network sizes. %
Various works modeled delay bounds as an \ac{ILP} for optimizing routes.
\cite{DeDinechin2014} used a linear formulation of the \ac{NC} end-to-end delay bound for optimizing the network-on-chip of a many core processor.
Their approach showed promising results compared to a nonlinear formulation on a small network. %
Similarly, \cite{Schweissguth2020} recently proposed another \ac{ILP} formulation tailored to \ac{TSN} and multicast flows, optimizing paths and schedules.
For both \cite{DeDinechin2014} and \cite{Schweissguth2020}, the scalability of both approaches to larger networks remains unclear.

In the scope of \ac{TSN}, \cite{Laursen2016} applied worst-case delay calculations in combination with a greedy optimization approach.
While the results show improvements over a shortest path approach, the formulation is tailored to the \ac{TSN} schedulers and the optimality of the solution is difficult to assess.
More recently, \cite{Bulbul2022} applied reinforcement learning to routing for routing for \ac{TSN} to meet flow deadlines.
Their method is based on packet-level simulation of the network, effectively giving no guarantees about the network delays computed.

\subsubsection*{Derivation of service requirements}

An \ac{NC}-based approach that can derive a lower bound on the system service was proposed in~\cite{Vastag11,Buchholz2017}.
It extends the (min,plus)-algebraic \ac{NC} with novel theory to take as input a function upper bounding the delay to be guaranteed under a certain load level.
However, the approach is currently restricted to \ac{FIFO} systems.
\ac{DiffNC} can perform the same task, yet without any restricting assumptions. 
It can fully use existing algebraic \ac{NC} theory and analyses.

\subsubsection*{NC combined with other methods}

(min,plus) algebra can be replaced with (max,plus) to better fit discrete event systems \cite{Liebeherr2017}.
\ac{NC} was paired with event stream theory \cite{Boyer2016} and with timed automata \cite{Lampka2009} for state-based system modeling.
\ac{NC} has been applied to the component-based models of real-time systems \cite{Thiele2000}, giving rise to the so-called real-time calculus.

Various formulations of the (min,plus) algebra as \ac{LP} were proposed, either addressing networks without assumptions on the multiplexing of flows \cite{Bouillard2010}, or with \ac{FIFO} scheduling \cite{Bouillard2015}.
These formulations provide tight delay bounds but scale poorly, as shown by \cite{Bondorf2017a} and later also in \cref{sec:optlp}.
These concerns were partially addressed recently in \cite{Bouillard2022}.
Additionally, \cite{Dang2014} also proposed an \ac{ILP} for optimizing \ac{TDMA} schedules in combination with \ac{NC}.

Finally, \ac{ML} was recently brought to \ac{NC} to speed-up costly network analyses originally requiring a search mimicking optimization in the algebraic approach.
DeepTMA was proposed in \cite{Geyer2019b,Geyer2020b} as a framework for predicting the best contention model.
Similarly, DeepFP \cite{Geyer2021a} targeted the prediction of best flow prolongation.
\cite{Mai2021} applied similar deep learning techniques for checking feasibility of network configurations, yet not for their synthesis.

%% file: netcal_background.tex
\section{Deterministic Network Calculus}
\label{sec:dncintro}

\acf{DNC} is built around two concepts \cite{LeBoudec2001}:
\begin{enumerate*}[label=\textit{\alph*)}] 
	\item the network of deterministic queueing locations crossed by constrained data flows, and
	\item resource modeling with cumulative functions in interval time.
\end{enumerate*}
Based on a fully specified model, a traditional \ac{DNC} analysis derives an upper bound on an analyzed flow's end-to-end delay.

We are interested in networks with point-to-point connections between devices, such as IEEE Ethernet and its extensions \ac{TSN} and \ac{AFDX}.
A model of such a network usually consists of the devices, e.g., switches, that are connected by undirected links.
I.e., an undirected graph $\mathcal{G}_{\text{device}} = (\mathcal{D}, \mathcal{L})$ is formed from devices $d\in\mathcal{D}$ and links $l\in\mathcal{L}$.
Most relevant for the \ac{NC} queueing analysis is the behavior at the devices' output ports.
Given the point-to-point connection, we therefore split the undirected links into directed ones such that there is one queueing location per directed link.
For convenience, we create the edge-to-vertex dual of this graph such that these queueing locations are represented by vertices.
The behavior at queueing locations can be defined by a wide range of scheduling policies, ranging from single \ac{FIFO} queues, over round robin schedulers (e.g., WRR, DRR) to a combination priority queues, shaping, time-triggered scheduling etc. as, for example, defined by \ac{TSN}.
For schedulers with static configurations to seperate service provision, the respective vertex is split into the defined number of priority levels or round robin class etc.
We call these vertices the servers and the resulting graph the server graph $\mathcal{G} = (\mathcal{S}, \mathcal{E})$ where each server $s\in\mathcal{S}$ forwards queued data.
The guaranteed service is expressed by non-negative, non-decreasing functions
$\mathcal{F}^{+}_{0}\!=\!\left\{ f:\mathbb{R}^{+}\rightarrow\mathbb{R}^{+}\,|\,f(0)\!=\!0,\;\forall s\le t\,:\,f(t)\!\geq\!f(s)\right\}$
called service curves.
\begin{definition}[Service Curve]
	If a server receives a data input $A\in\mathcal{F}^{+}_{0}$ and produces an output $A'\in\mathcal{F}^{+}_{0}$,
	then it is said to offer service curve $\beta \in \mathcal{F}_0$ iff
	\begin{equation}
	\label{eq:service_curve}
	\forall t : A'(t) \geq \inf_{0 \leq d \leq t} \{ A(t - d) + \beta(d) \}.
	\end{equation}
\end{definition}
The \ac{NC} literature provides a rich set of results to derive the service curves for different schedulers' service classes, e.g., \ac{TSN} \cite{Zhao2018,Zhao2020}.

Data flows cross the graph~$\mathcal{G}$ from a source server to a given number of destination servers, both in~$\mathcal{S}$.
The unicast or multicast flow path/s are subgraphs of $\mathcal{G}$ and a restriction on the data sent by a flow is assumed to be known at its source.
\begin{definition}[Arrival Curve]
	\label{def:Arrival-Curve}
	Given a flow $f$ described by $A\in\mathcal{F}^{+}_{0}$, 
	a function $\alpha\in\mathcal{F}_{0}$ is an arrival curve for $f$ iff
	\begin{equation}
	\label{eq:arrival_curve}
	\forall\,0\leq d\le t\,:\,A(t)-A(t-d)\leq\alpha(d).
	\end{equation}
\end{definition}

A traditional \ac{NC} analysis takes as input a fully specified queueing network model, i.e., a server graph with its service curves as well as flows, their paths and arrival curves.
It aims at deriving the end-to-end delay bound of a specific flow of interest (foi).
To do so, it needs to derive worst-case bounds on the mutual impact of flows at shared servers -- the queueing aspects not quantified by the model itself.
Our work is based on traditional (min,plus)-algebraic \ac{NC} analyses that assume the absence of cyclic dependencies between flows as well as the arbitrary multiplexing assumption.
These analyses derive a (min,plus)-algebraic term that bounds the end-to-end delay, consisting of the following operations:
\begin{definition}[Operations]%
	\label{def:MinPlusOperations}
	Given $\beta_1,\beta_2,\alpha_1,\alpha_2$, we can
	\begin{eqnarray}
	\text{aggregate:} &  & \hspace{-6.5mm}\left(\alpha_1+\alpha_2\right)\left(d\right) = \alpha_1\left(d\right)+\alpha_2\left(d\right)\!,\\
	\text{concatenate:} &  & \hspace{-6.5mm}\left(\beta_1\otimes \beta_2\right)(d) = {\displaystyle \inf_{0\leq u\leq d}}\left\{ \beta_1(d-u)+\beta_2(u)\right\}\!,\\
	\hspace{-5.5mm}\text{output bound:} &  & \hspace{-6.5mm}\left(\alpha_1\oslash \beta_1\right)(d) = \sup_{u\geq0}\left\{ \alpha_1(d+u)-\beta_1(u)\right\}\!,\\
	\text{left-over:} &  & \hspace{-6.5mm}\left(\beta_1\ominus \alpha_1\right)(d) = \sup_{0\leq u \leq d} \left\{\beta_1(u) - \alpha_1(u)\right\}\!, \\
	\text{delay bound:} &  & \hspace{-6.5mm} h(\alpha_1 ,\beta_1) = \inf\left\{ d\geq0|\!\left(\alpha_1\oslash\beta_1\right)(-d)\leq0\right\}\!\hspace{4mm}
	\end{eqnarray}
	flows and servers, respectively, in a worst-case manner.
\end{definition}

In this article, we will apply the \ac{SFA} \cite{LeBoudec2001} as the set of rules to derive the (min,plus)-algebraic, delay-bounding term from the server graph. %

%% file: diffnc.tex
\section{Differential Network Calculus}
\label{sec:diffnc}

The steps towards computing a delay bound as described in \Cref{sec:dncintro} cater to an analysis that requires a fully specified model.
I.e., in case there are design alternatives, each of these needs to be fully specified and analyzed independently.
A related topic are \ac{NC} analyses that explore alternative orders of (min,plus)-operations internally.
There, exhaustive enumeration may be feasible \cite{Bondorf2017a} but is often prohibitive \cite{Bondorf2017c}.
For either case, it was shown to be possible to design heuristics that vastly increase efficiency at little cost in terms of loss of delay bound accuracy \cite{Geyer2019b,2023-GSB-1}.
We propose \acl{DiffNC} (\acs{DiffNC}) that allows applying gradient-based \ac{NLP} to efficiently find a design alternative in a design space, subject to \ac{NC} delay bounds and at very low cost loss of accuracy.

\subsection{Generalized, parameterized \ac{NC} model}
\label{sec:formulation}

We generalize the server graph model of \ac{NC} to a more comprehensive, parameterized one where newly added parameters define design space alternatives.
We simply allow each variable in the model (e.g., those defining an arrival or service curve) to remain open and each part of the \ac{NC} model (e.g., entire curves) to be accompanied by an open parameter.
The added parameters can, e.g., be binary to decide whether the parameterized part of the model is to be considered;
continuous parameters can be added for weighing parameterized parts.
Combinations of parameters can also be restricted to express certain mutually exclusive design alternatives.
Without further restrictions on the parameters or their combinations, finding a parameter setting may become a \ac{MINLP}.
In this article, we exemplarily consider two instances of the generalized, parameterized \ac{NC} model.

\subsubsection{Alternative flow paths}
\label{sec:path_assignment}

Let $\mathcal{G} = (\mathcal{S}, \mathcal{E})$ be the server graph of the analyzed communication network and let $\mathcal{F}$ be the set of flows crossing $\mathcal{G}$.
We adopt here a path flow model, where each flow $f_i\in\mathcal{F}$ is associated with multiple alternative paths $\mathcal{P}_{f_i}$.
We amend each potential path $j \in \mathcal{P}_{f_i}$, or rather the data sent over $j$ by $f_i$, with a binary variable $p_{f_{i,j}}$ such that \cref{eq:arrival_curve} becomes:
\begin{equation} \label{eq:vitual_arrival_curve}
	\forall\,0\leq d\le t\,:\, A_{f_{i,j}}(t) - A_{f_{i,j}}(t-d) \leq \alpha_{f_i}(d) \cdot p_{f_{i,j}}
\end{equation}

To express that only a single path can be taken by a flow, we add the following constraint to our model:
\begin{equation} \label{eq:p_ij_sum}
	\sum_{j \in \mathcal{P}_{f_i}} p_{f_{i,j}} = 1, \forall f_i \in \mathcal{F}
\end{equation}

\begin{figure}[h!]
	\centering
	\includegraphics[width=.75\columnwidth]{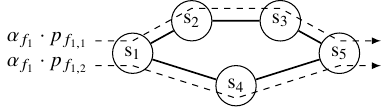}
	\caption{Illustration of virtual flow concept with one flow taking two potential paths in the server graph.}
	\label{fig:illustration_virtual_flow}
\end{figure}

We call the alternatives along the different potential paths \emph{virtual flows}.
Virtual flow arrival curves would then be set to $0$ on the non-optimal paths, and remain constrained by $\alpha_{f_i}$ on the optimal paths.
See \cref{fig:illustration_virtual_flow} for an illustration of the model.

Due to our formulation, servers may be overloaded, leading to invalid network configurations.
Hence the following constraint is required for each server $s$ in the server graph:
\begin{equation}
	\sum_{f \in \mathcal{F}_s} \mathit{rate}_f \leq \mathit{rate}_s
\end{equation}
with $\mathcal{F}_s$ the set of flows traversing server $s$.

As expected, we essentially defined a \ac{MINLP}, a combinatorial optimization problem with a number of potential solutions growing in $\mathcal{O}(|\mathcal{F}|^{|\mathcal{P}|})$.

\subsubsection{Flow priority assignment}

\begin{figure}[h!]
	\centering
	\includegraphics[width=.6\columnwidth]{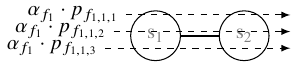}
	\caption{Virtual flow concept for priority assignment.}
	\label{fig:illustration_virtual_priority}
\end{figure}

We can use the virtual flow concept to define the search space for (network-wide) priority assignment.
Namely, a flow $f_i$ is extended to a set of virtual flows, one for each priority class.
For each virtual flow of $f_i$ with its potential path $j$ and potential priority class $k$, we define $p_{f_{i,j,k}}$ as a binary variable representing the choice of path $j$ and priority $k$ for flow $f_i$.
The arrival curve of each virtual flow is then amended by $p_{f_{i,j,k}}$, analog to \cref{eq:vitual_arrival_curve}. 
Again, the sum of $p_{f_{i,j,k}}$ variable must be equal to $1$ to enforce that only one virtual flow is selected, i.e., only one path and priority class are globally defined per flow.
\Cref{fig:illustration_virtual_priority} illustrates this model.

\subsection{Generalized \ac{NC} analysis by constrained \ac{NLP}}
\label{sec:constr_nlp}

Traditional \ac{NC} analyses such as \ac{SFA} would not be able to capture the virtual flow model.
Their backtracking of flow dependencies can succeed by ignoring the newly introduced binary variables, 
i.e, by including all alternatives of a virtual flow in the resulting (min,plus)-algebraic \ac{NC} term.
Analysis results are then valid, yet, overly pessimistic.
For an example of this practice, see \cite{2016-BG-1} where multicast flows are converted into a set of partially overlapping unicast flows.
If we extended the resulting (min,plus)-algebraic \ac{NC} term with the binary variables, then any unchanged traditional analyses would not be able to analyze this term anymore.
To solve this problem, we propose to use nonlinear optimization as analysis method.

\subsubsection{Closed-form expressions}

First, we need to derive closed-form expressions of the \ac{NC} operations presented in \cref{def:MinPlusOperations}.
We restrict our models to the curves shapes most commonly used in practice for industrial networks:
the rate-latency service curve $\beta_{R,L}$ and the token-bucket arrival curve $\gamma_{r,B}$:
\begin{eqnarray}
	\beta_{R,L}(t) & = & R [t - L]^+, \forall t \geq 0 \\
	\gamma_{r,B}(t) & = & B + r \cdot t, \forall t \geq 0
\end{eqnarray}
with $R$, $L$, $r$, and $B$ in $\R^+$, and $[x]^+ = x$ if $x \geq 0$ and 0 otherwise.

Applying \ac{NC}'s (min,plus)-algebraic operations to the above curve shapes, the following closed-form expression can be derived:
\begin{lemma}[Closed-form expression of \ac{NC} operations]
	\label{thm:nc_ops}
	With the assumption of using rate-latency service curves and token-bucket arrival curves, the \ac{NC} operations listed in \cref{def:MinPlusOperations} have the following closed-form solutions:
	\begin{eqnarray}
		\text{aggregation:} & \gamma_{r_1,B_1} + \gamma_{r_2,B_2} = \gamma_{r_1+r_2,B_1+B_2} \\
		\text{concatenation:} & \beta_{R_1,L_1} \otimes \beta_{R_2,L_2} = \beta_{\min(R_1, R_2),L_1 + L_2} \\
		\text{output bounding:} & \gamma_{r,B} \oslash \beta_{R,L} = \gamma_{r,B+r \cdot L} \\
		\text{left-over:} & \beta_{R,L} \ominus \gamma_{r,B} = \beta_{R - r, (B + R \cdot L) / (R - r)} \\
		\text{delay bound:} & h(\gamma_{r,B}, \beta_{R,L}) = B / R + L
	\end{eqnarray}
	under the previously mentioned stability condition that $r < R$.
\end{lemma}

Note that we restrict the curve shapes for brevity, our approach is not limited to these curve shapes.
Using different shapes requires to provide the closed-form expressions as in \cref{thm:nc_ops}.
This can be relatively simple, e.g., for concave arrival curves and convex service curves.

\subsubsection{Constrained \ac{NLP} Modeling}

Next, we show here how to model the network design problem as a differentiable \ac{NLP} of the following form:
\begin{align}
\min_{x \in \R^n}      & \quad f(x) \label{eq:general_nonlin_obj} \\
\text{s.t.} & \quad  g_l \leq g(x) \leq g_u \label{eq:general_nonlin_constr}
\end{align}
with $f()$ and $g()$ differentiable functions w.r.t. $x$, and $g_l$ and $g_u$ the upper and lower bounds for $g()$.

To make this problem solvable in polynomial time, we apply a commonly used technique known as relaxation, namely: the $p_{f_{i,j}}$ binary variables are relaxed as continuous variables on the interval $[0, 1]$.
Following \cref{thm:nc_diff}, the end-to-end delay bound expression of a virtual flow is then differentiable w.r.t. the $p_{f_{i,j}}$ variables.
This relaxation technique transforms the \ac{MINLP} into a continuous \ac{NLP}, enabling the use of \ac{NLP} methods based on gradient information.

Using on the previous formulations, we define the following constrained nonlinear optimization problem that minimizes the average delay bound:
\begin{align} \label{eq:nonlin_obj}
	\min_{p_{f_{i,j}}, \forall f_i \in \mathcal{F}, j \in \mathcal{P}_{f_i}} & \quad \frac{1}{|\mathcal{F}|} \sum_{i, j} \textit{delay bound}(f_{i,j}) \cdot p_{f_{i,j}} \\
	\text{s.t.} & \quad 0 \leq p_{f_{i,j}} \leq 1, \forall f_i \in \mathcal{F}, j \in \mathcal{P}_{f_i} \\
	& \quad \sum_{j \in \mathcal{P}_{f_i}} p_{f_{i,j}} = 1, \forall f_i \in \mathcal{F} \\
	& \quad \sum_{i \in T(k)} r_i \cdot p_{f_{i,j}} \leq R_k, \forall k \in \mathcal{S} \label{eq:cstr_server_overload}
\end{align}
with $T(k)$ the set of virtual flows traversing server $k$ with service curve $\beta_{R_k,L_k}$, and $\textit{delay bound}(f_{i,j})$ the end-to-end delay bound of the virtual flow $f_{i,j}$ computed with any of the classical algebraic \ac{NC} analyses (e.g., \ac{SFA}).
Note, that we thus also generalized the analysis to consider the delay bounds of multiple flows $f_{i,j}$ simultaneously
whereas the classical analyses can only analyze a single flow of interest in isolation.
Due to the operations in \cref{thm:nc_ops}, \cref{eq:nonlin_obj} is a non-convex objective function.

\begin{figure*}[t!]
	\centering
	\includegraphics[width=.85\textwidth]{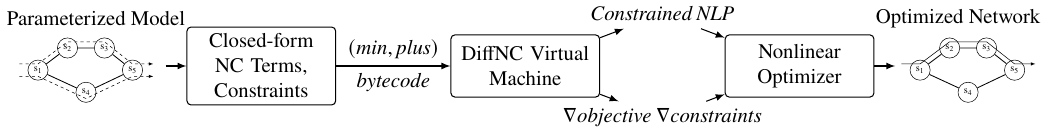}
	\caption{Illustration of the conceptual \ac{DiffNC} proceedings. First, the required (min,plus) operations are prepared as (min,plus) bytecode for our VM. The precomputed operations are then used during the iterative optimization process to avoid unnecessarily computing them at each optimization iteration. The gradient is computed using a backward pass on the (min,plus) operations.}
	\label{fig:system_overview}
\end{figure*}

\subsubsection{Extended Analysis Capabilities}

We already showed the ability to consider multiple flows' delay bounds simultaneously.
With our formulation, we also enable a wider range of constraints and objective functions.
Constraints can be added such that a maximum delay requirement is satisfied for a given flow, saving us a subsequent check against the requirement:
\begin{equation} \label{eq:delay_req}
	\sum_{j \in \mathcal{P}_{f_i}} \textit{delay bound}(f_{i,j}) \cdot p_{f_{i,j}} \leq \textit{requirement}
\end{equation}

This formulation is able to express complex objectives, enabling finer control over the type of solution which is required.
A popular mathematical framework for describing hard or soft requirements on network performance (such as delay or bandwidth) is the concept of utility-based network optimization introduced in \cite{Kelly1998}.
The objective function can be formulated with nonlinear utility functions $U_i$ for the delay bounds:
\begin{equation}
	\min_{p_{f_{i,j}}, \forall i, j} \sum_{i} U_i \left(\sum_j \textit{delay bound}(f_{i,j}) \cdot p_{f_{i,j}} \right)
\end{equation}
with $U_i$ a differentiable utility function mapping the delay bonds to a utility value in the interval $[0, 1]$.

Additionally, aspects such as the tail of the delay bound distribution can be minimized by defining the objective function:
\begin{equation}
	\min_{p_{f_{i,j}}, \forall i, j} \max_{i} \left(\sum_j \textit{delay bound}(f_{i,j}) \cdot p_{f_{i,j}} \right)
\end{equation}

The optimization formulation can be applied to any curve parameter, including the service curve parameters.
This means that the optimization formulation can also be defined w.r.t. scheduler characteristics.

\subsection{Differentiation of expressions for gradient-based \ac{NLP}}
\label{sec:diff_delay_term}

\ac{NLP} techniques based on gradient information -- such as Newton's method -- are known to usually outperform other \ac{NLP} optimization techniques.
Therefore, we show here that the delay bounding terms as well as the constraints are differentiable and confirm later in \cref{sec:numerical_evaluation} that this is key to a high performance analysis.

From \cref{thm:nc_ops}, the following theorem is derived:
\begin{theorem}[Differentiability of delay expression]
	\label{thm:nc_diff}
	With the assumption of using rate-latency service curves and token-bucket arrival curves, a \ac{NC} end-to-end delay bound is differentiable w.r.t. the curves parameters.
\end{theorem}

\textit{Proof:} Using the closed-form (min,plus) operations from \cref{thm:nc_ops}, all \ac{NC} operations use the following basic operators: addition, multiplication, division and min.
For the $\min$ operator, we use the following partial derivates for $x \neq y$:
\begin{align*}
	\frac{\partial \min(x, y)}{\partial x} = \begin{cases}
		1 \text{, if } x < y \\
		0 \text{, if } x > y
	\end{cases} & & \frac{\partial \min(x, y)}{\partial y} = \begin{cases}
	0 \text{, if } x < y \\
	1 \text{, if } x > y
	\end{cases}
\end{align*}
All of the applied operators are then differentiable, proving \cref{thm:nc_diff}.
\hfill $\square$

Partial derivates for the min operator used in the above proof are easily implemented using the Heaviside step function.

Following the previous theorems, \crefrange{eq:nonlin_obj}{eq:cstr_server_overload} are differentiable w.r.t the relaxed $p_{f_{i,j}}$ variables.

\section{Automatic differentiation and optimization} %
\label{sec:diffnc_ad}

The previous theorems build the mathematical foundations of \ac{DiffNC}.
Connecting them to the conceptual \ac{DiffNC} proceeding is straight-forward.
\Cref{fig:system_overview} illustrates the steps.
We detail here how to put \ac{DiffNC} into practice in order to efficiently compute partial derivatives of the end-to-end delay bounds w.r.t. the curves parameters.
We also present a preliminary numerical evaluation of the performances of our toolchain.

\subsection{Software architecture}

While computer-assisted symbolic differentiation could be used for deriving closed-form expressions of the gradient, our initial numerical evaluations with SymPy \cite{Meurer2017} showed that this method had difficulties scaling to networks with 100+ flows.

To overcome this scalability issues, we selected \ac{AD}.
It is a family of techniques based on the calculus' chain rule for efficiently and accurately evaluating derivatives of numeric functions expressed as computer programs.
This technique has gained a lot of popularity recently due its wide use in computing packages used for machine learning \cite{Baydin2018}.

While the first version of \ac{DiffNC} from \cite{GeyerBondorf_INFOCOM2022} was based on CasADi \cite{Andersson2019}, we implemented our own \acf{AD} tool in Go for this work, specialized for (min,plus) operations and delay bound calculations.
As shown later in \cref{sec:eval:toolchain,sec:eval:execution_time}, our implementation enables us better scalability on large networks and to parallelize various parts of the code.
This stems from the fact that most available \ac{AD} tools are targeting relatively small computation graphs (i.e. series of mathematical operations described as a directed graph) with arithmetically intense operations such as large matrix multiplications.
This specialization makes these tools poorly scalable for \ac{NC} operations, as millions of (min,plus) operations are required to compute delay bounds in our use-cases.
Our tool also directly uses the (min,plus) operations, enabling a better scalability compared to a conversion to basic mathematical operations.

As network analysis, we choose \ac{SFA} and execute its backtracking to derive a (min,plus)-algebraic \ac{NC} term.
Then we extend the term with the $p_{f_{i,j}}$ binary variables (see \cref{sec:formulation}) and convert it to closed form expressions in (plus,times) algebra according to \cref{thm:nc_ops}.
Combined with nonlinear optimization methods using gradients -- as detailed later in \cref{sec:diffnc_implementation} -- our generalized network models can efficiently be optimized.

Additionally, since most optimization methods require multiple evaluations of the objective function, our tool also allows us to run the \ac{NC} network analysis a single time and generate a so-called computation graph.
This graph translates the delay expressions (i.e., the objective function) to a combination of basic mathematical operations (addition, multiplication, etc.) and (min,plus) operations.
These saved operations can then be evaluated multiple times in our so-called \emph{\ac{DiffNC} \ac{VM}}, without requiring to run the \ac{NC}-specific parts again.

A network to-be-optimized with alternative flow paths is used as input of our framework.
Based on the end-to-end delay bounds calculations, the computation graph of the objective function and its gradient, and the constraint functions and its gradients, are then generated and compiled for our \ac{DiffNC} \ac{VM}.
The compiled formulas are then used as input to a nonlinear optimizer supporting the generalized \ac{NLP} presented in \cref{eq:general_nonlin_obj,eq:general_nonlin_constr}.

\subsection{The Frank-Wolfe algorithm}
\label{sec:fw}

We detail here the nonlinear optimization part using gradient-based constrained optimization methods with open-source implementations.

We selected the Frank-Wolfe algorithm \cite{Frank1956,Braun2022} -- also known as conditional gradient method -- as our main solution for solving the \ac{NLP} described in \cref{sec:diffnc}.
To optimize a \ac{NLP} with objective function $f$, the Frank-Wolfe algorithm can be summarized as the following loop.
Given a solution $x_k$ at iteration $k$:
\begin{itemize}
	\item \emph{Step 1:} Find the solution $s_k$ to the linearized version of the \ac{NLP} using the gradient information (i.e., $s_k^T \triangledown f(x_k)$),
	\item \emph{Step 2:} Update $x_{k+1} \gets x_k + \delta (s_k - x_k)$.
\end{itemize}
The choice of $\delta$ determines the so-called step size of the Frank-Wolfe algorithm.

While this algorithm was designed to solve convex \ac{NLP}, it was shown to also perform well on non-convex problems in practice by choosing $\delta = 1 / \sqrt{k+1}$ \cite{Braun2022}.
As we show later, it achieves good optimality at a low computational cost.

Several extensions of this algorithm have been proposed in the literature.
We explored some of them and found that its variant with momentum \cite{Braun2022} achieves good results in our problem setting.
Momentum is a technique to build inertia in a direction in the search space in order to overcome the oscillations of noisy gradients.

We implemented the standard Frank-Wolfe algorithm to be our main optimization algorithm in the following evaluations.
\Cref{sec:eval:frankwolfe} provides a peak into the performance of the Frank-Wolfe algorithm with momentum.

\subsection{Implementation details of our toolchain}
\label{sec:eval:toolchain}

We describe and numerically evaluate in this section key elements of our toolchain to enable fast evaluations.
The numerical evaluations were performed on the datasets described later in \cref{sec:dataset}.

\subsubsection{Parallelization}
\label{sec:eval:parallelization}

We evaluate in this section the impact of parallelization of the operations on the \ac{DiffNC} \ac{VM}.
For our implementation, we use different strategies for parallelization.
First, we parallelize the different delay bound computations both during the preparation of the (min,plus) terms and during the execution in the \ac{DiffNC} \ac{VM}.
These computations are independent of each other and our objective function from \cref{eq:nonlin_obj} requires the computation of the delay bounds of all flows in the network.
Secondly, we also parallelized part of the computations during the preparation of the (min,plus) terms of the \ac{SFA} network analysis.

\begin{figure}[h!]
	\centering
	\includegraphics[width=\columnwidth]{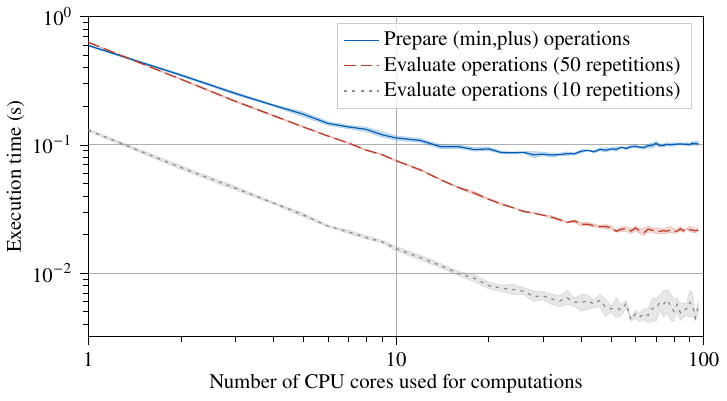}
	\caption{Impact of parallelization on the \ac{DiffNC} \ac{VM} for the delay bounds calculation of the \ac{AFDX} network. Execution times were measured on an AMD EPYC 7702P.}
	\label{fig:execution_time_parallelization_afdx}
\end{figure}

The impact of the parallelization is highlighted in \cref{fig:execution_time_parallelization_afdx}, where the delay bound calculations of all flows from the \ac{AFDX} network were performed.
We notice that, as we increase the number of cores used for the computations, the total execution time of the analysis is reduced.
Overall, a gain of more than one order of magnitude was possible thanks to parallelization.

The benefit of precomputing the (min,plus) computations is also illustrated in \cref{fig:execution_time_parallelization_afdx}.
Running the compiled analysis can be performed 50 times faster than running it without precomputation.

\subsubsection{Impact of VM vs. native code}

\begin{figure}[h!]
	\centering
	\includegraphics[width=\columnwidth]{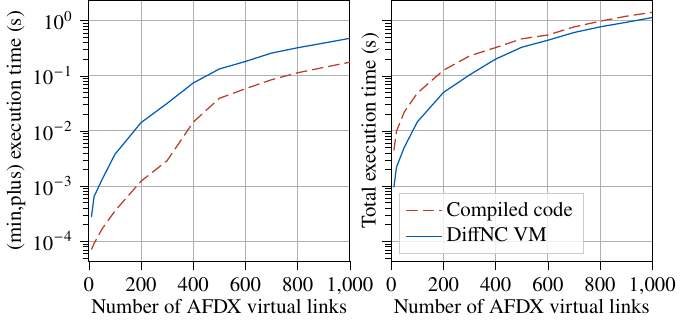}
	\caption{Comparison of execution time of \ac{DiffNC} VM against compiled code on the \ac{AFDX} network. Only one core is used for this benchmark.}
	\label{fig:execution_time_comparison_netcalvm_cjit_afdx}
\end{figure}

Finally, we also benchmark our \ac{DiffNC} \ac{VM} against compiled code, i.e., against the same \ac{NC} operations compiled to Intel assembly.
Results on the \ac{AFDX} topology are presented in \cref{fig:execution_time_comparison_netcalvm_cjit_afdx}.

When comparing the execution time of the (min,plus) operations, the \ac{DiffNC} \ac{VM} is slower than compiled code.
Yet, the cost of compiling the (min,plus) operations to assembly makes it overall slower than using the \ac{DiffNC} \ac{VM}, as shown by the total execution time.

%% file: linear_formulation.tex
\section{Non-algebraic \ac{NC} alternative}
\label{sec:optlp}

\ac{DiffNC} is not the first attempt at combining optimization techniques with \ac{NC}.
An \ac{LP} formulation of an \ac{NC} model was proposed in \cite{Bouillard2010}.
It converts the equations introduced in \cref{sec:dncintro} to linear constraints, under the assumption that curves are piecewise linear functions, either concave or convex. %
Flows are backtracked to derive constraints capturing, among others, their mutual impact.
Complexity grows exponentially when computing a flow's tight delay bound and a heuristic called \ac{ULP} was proposed.
The \ac{ULP} was shown to have only limited scalability~\cite{Bondorf2017a}, yet it is our only hope for a non-algebraic \ac{NC} competitor.
We complement the \ac{ULP} to include our $p_{f_i,j}$ variables and to optimize for multiple flows.
The resulting formulation is able to find configurations, yet preliminary evaluation already shows that it scales poorly.

The \ac{ULP} is based around two classes of variables.
Time variables $t_h \in \R^+$ represent departure or arrival time of bits of data of the flows at the different servers of the network.
Function variables $A_{f_i}^{s_k}(t_h) \in \R^+$ represent the departure and arrival processes of the data of flows at the different servers of the network, i.e., the arrival and departure functions introduced in \cref{sec:dncintro} as $A(t)$ and $A'(t)$.

Based on these variables, the \ac{ULP} translates arrival curves from \cref{eq:arrival_curve} as linear constraints:
\begin{equation} \label{eq:arrival_curve:constraint}
	A_{f_i}^{s_k}(t_{h+1}) - A_{f_i}^{s_k}(t_h) \leq \alpha_i(t_{h+1} - t_h), \forall s_k, f_i
\end{equation}
and similarly service curves from \cref{eq:service_curve} as:
\begin{equation}
	\sum_{f_i} \left( A_{f_i}^{s_k}(t_{h+1}) - A_{f_i}^{s_k}(t_h) \right) \geq \beta_k(t_{h+1} - t_h), \forall s_k
\end{equation}
Additional constraints representing, e.g., causality are also added.
We refer to \cite{Bouillard2010} for a full formulation.%

We extend here this formulation to take into account different paths for the flows.
For each flow $i$ and each potential path $j$, we define the variables $A_{f_{i,j}}^{s_k}(t_h)$ as the departure and arrival processes of the data of the virtual flows along the path $j$.
The variables $A_{f_{i,j}}^{s_k}(t_h)$ are constrained as in the original formulation from \cite{Bouillard2010} as if they were normal flows.
Following \cref{eq:vitual_arrival_curve}, the following constraints are added:
\begin{equation} \label{eq:lp_As_constr}
	A_{f_{i,j}}^{s_k}(t_h) \leq M \cdot p_{f_{i,j}}, \forall s_k, f_{i,j}, t_h
\end{equation}
with $M$ a large constant chosen such that $\alpha_i(t_{h+1} - t_h) \leq M, \forall t_{h+1}, t_h$ in the \ac{LP} formulation.
Using the big-M method, \cref{eq:lp_As_constr} achieves the same effect as \cref{eq:vitual_arrival_curve}: the $A_{f_{i,j}}^{s_k}(t_h)$ are constrained to 0 on the paths where $p_{f_{i,j}} = 0$ -- i.e., removing their impact on the delay calculation of the other flows -- and leaving them unconstrained when $p_{f_{i,j}} = 1$.

While this \ac{ULP} formulation of the optimal routing problem is attractive, it suffers from two important drawbacks: difficulty for expressing delay constraints and optimization goals, and poor scalability.
The first drawback of this approach is that some requirements regarding the optimization problem are not straightforward to translate into the \ac{ULP}.
This drawback mainly stems from the fact that the delay bound itself is calculated by maximizing an expression in the \ac{ULP}.
This leads to difficulty at implementing an objective function which would minimize average delay bounds.
Similarly, adding a constraint regarding a maximum delay requirement as in \cref{eq:delay_req} is not straightforward: the objective function maximizes the delay bound, but such a constraint would result in an underestimation of the delay bound in some cases.

\begin{figure}[h]
	\includegraphics[width=\columnwidth]{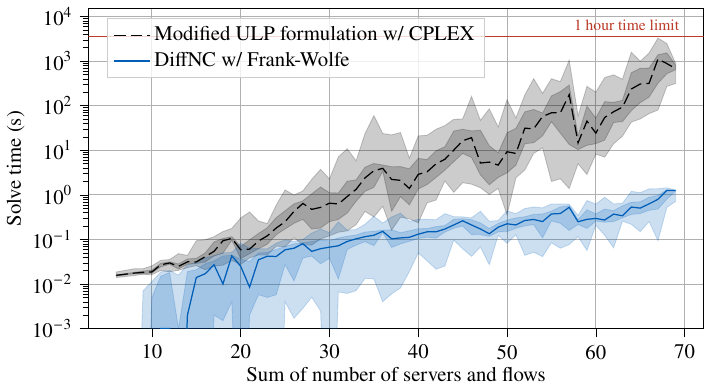}
	\caption{Solve time of the \ac{ULP} formulation against network size using CPLEX. The curves represent respectively the 10, 25, 50, 75 and 90 percentiles.}
	\label{fig:nclp_networksize_vs_solvetime}
\end{figure}

Secondly, as noted in \cite{Bouillard2010,Bouillard2014HDR} and subsequently numerically illustrated in \cite{Bondorf2017a}, the \ac{ULP} is only tractable on relatively small networks due to its exponentially growing number of constraints.
To illustrate this point, we evaluated our modified \ac{ULP} including the $p_{f_i,j}$ variables and the constraints from \cref{eq:lp_As_constr} on a set of randomly generated networks.
Details about the networks are explained later in \cref{sec:dataset}.
For the objective function, we maximize the sum of delay bounds.
We extended the \ac{ULP} implementation from NCorg DNC v2.6.2~\cite{Bondorf2014} for our evaluation.
\Cref{fig:nclp_networksize_vs_solvetime} illustrates the time to find a solution with a time limit of 1 hour using IBM's CPLEX 20.1.0 on an Intel Xeon Gold 5120 at \SI{2.2}{\GHz}.

As expected, the solve time grows exponentially, exceeding the one hour time limit even on small networks.
This result highlights why an alternative solution for optimizing networks is necessary for larger networks.
As a further motivating comparison, \cref{fig:nclp_networksize_vs_solvetime} also illustrates the optimization time on the same networks with our contributed approach: \ac{DiffNC} with Frank-Wolfe.%

%% file: evaluation.tex
\section{Numerical evaluation}
\label{sec:numerical_evaluation}

We evaluate in this section our approach on a wide range of networks.
We illustrate its scalability and compare it against other optimization methods.

\subsection{Competing nonlinear optimization algorithms for \ac{DiffNC}}
\label{sec:diffnc_implementation}

For our evaluation, we also selected \ac{SLSQP}.
It was shown in \cite{GeyerBondorf_INFOCOM2022} to perform the best in terms of optimality.
Finally, we used the \ac{MMA} \cite{Svanberg1987} and in conjunction with an augmented Lagrangian method \cite{Hestenes1969,Powell1969} in order to include the equality constraints from \cref{eq:p_ij_sum}.
For all these algorithms, the implementation from \cite{Johnson2020} is used.

Integer relaxation was used and the final solution is then converted back to integer and verified against the constraints.
Note that these are local optimization methods, each requiring a starting point.
For our evaluation and metrics, a single randomly generated starting point was used.

\subsection{Other heuristics}
\label{sec:otherheuristics}

To benchmark our approach against potential competitors, the following other optimization methods were selected.
They include both naïve and greedy approaches, as well as other optimization techniques often used for solving constrained combinatorial problems.
A maximum of 500 evaluations of the objective function has been defined for the heuristics described here.

\subsubsection{Randomized search}
In this greedy approach, $500$ random combinations of paths are chosen and evaluated.
The combination leading to the best objective is presented.
In the figures, this method is labeled as Random.

\subsubsection{Hop-count shortest path}
For this approach, the path minimizing the number of hops for each flow is selected.
This approach does not use other information about the network and is equivalent to a traditional Dijkstra shortest-path algorithm.

\subsubsection{Minimum-delay shortest path}
This approach is similar to the previous one, except that we partially take into account the arrival and service curves in the network.
The minimum delay bound that \iac{NC} end-to-end delay analysis can compute for a flow is in the absence of crosstraffic impact,
e.g., if the flow had highest priority among all flows. 
In that case, the end-to-end delay bound is $h(\gamma_{r,B},\beta_{R_j,L_j}) = L_j+\frac{B}{R_j}$ with $R_j$ the minimum rate on its path $j\in\mathcal{P}_i$ and $L_j$ the sum of server latencies.
We use this as a proxy metric and select the path with minimum end-to-end delay in hypothetical absence of crosstraffic.

\subsubsection{Non-gradient based methods}
We also evaluate the Nelder-Mead \cite{Nelder1965} and Subplex \cite{Rowan1990} algorithms, both direct search methods based on the simplex algorithm.
Both algorithms do not make use of the gradient information.

Our original work \cite{GeyerBondorf_INFOCOM2022} already evaluated other heuristics based on meta-heuristics or evolutionary algorithms.
They were omitted here since our previous evaluation showed that they underperformed compared to \ac{DiffNC}.

\subsection{Evaluated networks}
\label{sec:dataset}

To numerically evaluate our approach, we randomly generated a set of evaluation networks.
First, a random amount of servers was generated, connected in a directed graph.
Each server has a rate-latency service curve, with rate and latency parameters randomly sampled from a uniform distribution.
A random amount of source-destination pairs was then generated for flows, each with a token-bucket arrival curve, with rate and burst parameters randomly sampled from a uniform distribution.
For each pair, a set of virtual flows were generated according to the available paths in the directed graph.

The goal of the optimization is to find the best path and the best priority (high or low, network-wide) for a given flow.
For each network, the minimization of the average end-to-end delay bound of the flows is used as objective function, computed using multicast \ac{SFA} under the assumption of arbitrary multiplexing \cite{2016-BG-1}.

Overall, our dataset contains topologies with up to 1000 flows, matching the number of flows found in some industrial settings \cite{Boyer2012,TamasSelicean2015,Belliardi2018}.
\Cref{tab:dataset:full} contains statistics about the dataset.

\begin{table}[h!]
	\centering
	\caption{Statistics about the generated dataset.}
	\label{tab:dataset:full}
	\begin{tabular}{lrrrr}
		\toprule
		\textbf{Number of}    & \textbf{Min} & \textbf{Mean} & \textbf{Median} & \textbf{Max} \\ \midrule
		Servers               &            8 &         17.08 &              16 &           31 \\
		Flows                 &            5 &        170.67 &             164 &         1001 \\
		Virtual flows         &            9 &        355.22 &             343 &         1884 \\
		Path combinations     &    \TP{1.08} &    \TP{46.04} &      \TP{44.10} &  \TP{229.08} \\
		Path + priority comb. &    \TP{2.58} &    \TP{97.41} &      \TP{94.28} &  \TP{530.41} \\ \bottomrule
	\end{tabular}
\end{table}

Additionally for the numerical evaluation performed in \cref{sec:optlp,fig:nclp_networksize_vs_solvetime}, a dataset containing smaller networks was also generated using the same approach.
\Cref{tab:dataset:lpeval} contains relevant statistics about this additional dataset.

\begin{table}[h!]
	\centering
	\caption{Statistics about the networks used for the evaluations in \cref{sec:optlp,sec:eval:optimality}.}
	\label{tab:dataset:lpeval}
	\begin{tabular}{lrrrr}
		\toprule
		\textbf{Number of}    & \textbf{Min} & \textbf{Mean} & \textbf{Median} & \textbf{Max} \\ \midrule
		Servers               &            3 &          8.68 &               8 &           18 \\
		Flows                 &            3 &          9.70 &               9 &           21 \\
		Virtual flows         &            4 &         18.62 &              17 &           45 \\
		Path combinations     &    \TP{0.30} &     \TP{2.07} &       \TP{1.81} &    \TP{5.52} \\
		Path + priority comb. &    \TP{1.20} &    \TP{11.32} &       \TP{9.20} &   \TP{32.81} \\ \bottomrule
	\end{tabular}
\end{table}

This is the same dataset as used in \cite{GeyerBondorf_INFOCOM2022}\footnote{Online \texttt{https://github.com/fabgeyer/dataset-infocom2022}}, with the addition that we also optimize for the best priority for a given flow.

Finally, we also used the \ac{AFDX} network from the Airbus A350 as a representative industrial network for our evaluations in \cref{sec:evaluation_afdx,sec:eval:execution_time}.
This network contains approximately 1100 multicast flows with an average of 8 destinations per multicast flow.
In this industrial network, multicast paths are fixed such that we focus only on optimizing the (network-wide) priority of the flows.

\subsection{Reduction of delay bounds}
\label{sec:eval:gaptobest}

We evaluate here the solution of each optimization method presented in \cref{sec:otherheuristics} on our evaluation dataset.
We use here \cref{eq:nonlin_obj} as objective function, i.e., we minimize the average end-to-end delay bound in the networks, an objective function found in many other related works.

We first compare the optimization methods using the result of the hop-count shortest path approach as a baseline.
We use the relative gap of the objective function as our metric, namely:
\begin{equation}
\mathit{RelGapShortestPath}_\text{method} = \frac{\mathit{objective}_\text{method}}{\mathit{objective}_{\text{shortest path}}} - 1
\end{equation}
Since we aim at minimizing delay bounds, a negative value of the relative gap means that the evaluated optimization method achieved better results than simply using shortest path.

\begin{figure}[h!]
	\centering
	\includegraphics[width=\columnwidth]{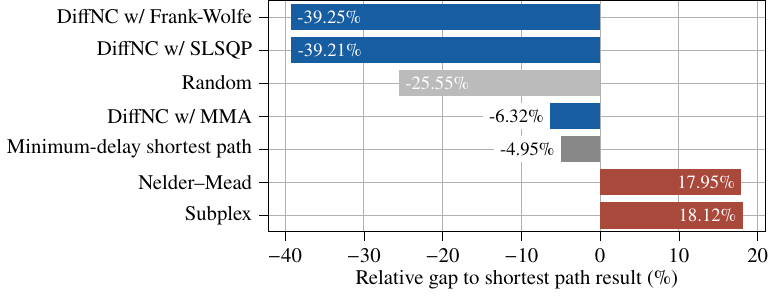}
	\caption{Average relative gap to the result of shortest path. Negative values mean optimizations outperform shortest path.}
	\label{fig:relative_gap_shortest_path}
\end{figure}

Results are presented in \cref{fig:relative_gap_shortest_path}.
\ac{DiffNC} with Frank-Wolfe is able to achieve the best results compared to all the other heuristics evaluated here, closely followed by \ac{SLSQP}.
Overall, Frank-Wolfe achieved a reduction of \SI{39.25}{\percent} of the average delay bounds.
Compared to the non-gradient-based algorithms, all gradient-based optimization methods based on \ac{DiffNC} achieve much better results.
Interestingly, the delay-based shortest path approach is able to surpass the non-gradient-based optimization methods, showing that a simple heuristic using domain knowledge about the model and analysis used in the optimization problem can be somewhat effective.

Given the large number of path combinations in some networks (larger than 10\textsuperscript{530} in some cases), the optimal network configurations are not known and cannot be computed in reasonable time by simply enumerating the combinations.
To address this, we use the best result which was obtained by any evaluated method as a baseline, called here virtual best.
We use the relative gap of the objective function to the best objective as metric:
\begin{equation}
\mathit{RelGapBest}_\text{method} = \frac{\mathit{objective}_\text{method}}{\mathit{objective}_{\text{virtual best}}} - 1
\end{equation}

\begin{figure}[h!]
	\centering
	\includegraphics[width=\columnwidth]{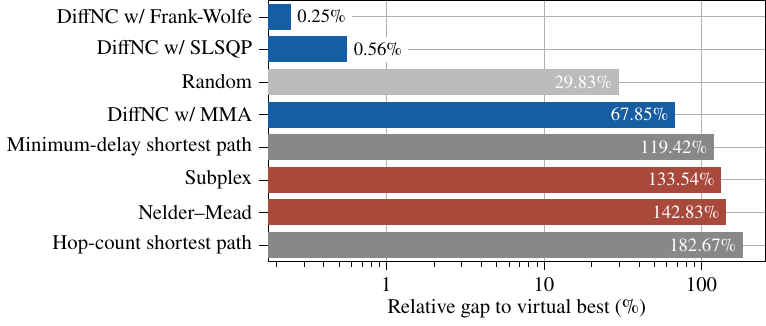}
	\caption{Average relative gap to the best objective. A value close to zero indicates a solution close to the best one.}
	\label{fig:relative_gap_best_objective}
\end{figure}

Results are presented in \cref{fig:relative_gap_best_objective}.
With an average relative gap of \SI{0.25}{\percent}, \ac{DiffNC} with Frank-Wolfe achieves the best results compared to all the other heuristics.
\ac{DiffNC} with \ac{SLSQP} closely follows it with an average relative gap of \SI{0.56}{\percent}.
Both methods outperform all the other heuristics by at least one order of magnitude.
Most observations made from \cref{fig:relative_gap_shortest_path} for the other methods also apply for \cref{fig:relative_gap_best_objective}.
The only noteworthy exception is that the \ac{NLP} algorithms not using gradient information are not beaten by the hop-count shortest path w.r.t. the relative gap to the virtually best method.

\subsection{Optimality gap}
\label{sec:eval:optimality}

We evaluate the gap of \ac{DiffNC} to the optimum found by exhaustive enumeration, restricted to networks from \cref{tab:dataset:lpeval} where the analysis terminates within \SI{1}{\hour}.
\Cref{tab:comparison_bruteforce} shows the results:
\ac{SLSQP} and Frank-Wolfe were able to respectively find the optimum in \SI{89.1}{\percent} and \SI{83.0}{\percent} of networks.
The relative gap to the optimum is of \SI{0.1}{\percent} and \SI{0.3}{\percent}, outperforming all the other methods by one or two orders of magnitude.

\begin{table}[h!]
	\centering
	\caption{Number of networks where the optimal solution from exhaustive enumeration was found and average relative gap to the optimal solution.}
	\label{tab:comparison_bruteforce}
	\begin{tabular}{lcc}
		\toprule
		\textbf{Method}             & \textbf{Opt. found} & \textbf{Rel. gap to opt} \\ \midrule
		DiffNC w/ SLSQP             & \SI{89.1}{\percent} &    \SI{0.1}{\percent}    \\
		DiffNC w/ Frank-Wolfe       & \SI{83.0}{\percent} &    \SI{0.3}{\percent}    \\
		Random                      & \SI{62.4}{\percent} &    \SI{3.9}{\percent}    \\
		DiffNC w/ MMA               & \SI{52.5}{\percent} &   \SI{29.8}{\percent}    \\
		Subplex                     & \SI{39.7}{\percent} &   \SI{36.5}{\percent}    \\
		Minimum-delay shortest path & \SI{36.7}{\percent} &   \SI{51.5}{\percent}    \\
		Hop-count shortest path     & \SI{30.3}{\percent} &   \SI{98.0}{\percent}    \\
		Nelder-Mead                 & \SI{25.3}{\percent} &   \SI{73.0}{\percent}    \\ \bottomrule
	\end{tabular}
\end{table}

\subsection{Application to an industrial network}
\label{sec:evaluation_afdx}

We evaluate in this section our approach on the \ac{AFDX} network from the Airbus A350.
We use the average reduction in delay bound compared to using only one priority level as our metric for evaluating \ac{DiffNC}, namely:
\begin{equation}
	\frac{\mathit{objective}_{n\:\mathit{priorities}}}{\mathit{objective}_{\mathit{one\:priority}}} - 1
\end{equation}

\begin{figure}[h!]
	\centering
	\includegraphics[width=\columnwidth]{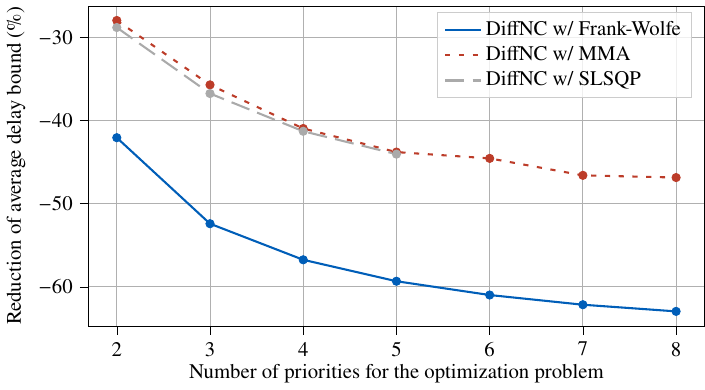}
	\caption{Reduction of delay bounds on \ac{AFDX} network compared to using only one priority in the network. Results for \ac{DiffNC} with \ac{SLSQP} with more than 5 priorities are omitted here since they take more than \SI{10}{\hour} to compute.}
	\label{fig:reduction_delay_bound_vs_nprios_a350_frankwolfe_vs_mma}
\end{figure}

Results are presented in \cref{fig:reduction_delay_bound_vs_nprios_a350_frankwolfe_vs_mma}.
\ac{DiffNC} with Frank-Wolfe is able to outperform both \ac{DiffNC} with \ac{MMA} and with \ac{SLSQP}.
This result confirms the conclusions from \cref{sec:eval:gaptobest} where a similar behavior was observed.

\subsection{Frank-Wolfe algorithm with momentum}
\label{sec:eval:frankwolfe}

Due to its good performances in terms of optimality and computational cost, the Frank-Wolfe algorithm has been extended in various works, as shown in a recent survey \cite{Braun2022}.
In order to explore the potential to further improve its performance in terms of optimality, we used its variant with momentum \cite{Braun2022}.%
In practice, momentum builds inertia in a direction in the search space and overcome the oscillations of noisy gradients.

\begin{figure}[h!]
	\centering
	\includegraphics[width=\columnwidth]{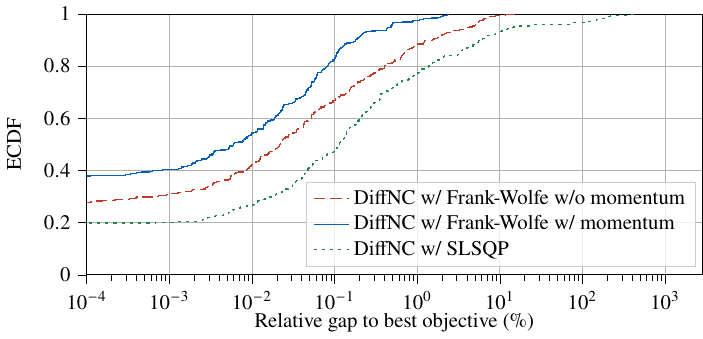}
	\caption{Impact of momentum on Frank-Wolfe optimality.}
	\label{fig:comparison_frankwolfe_momentum}
\end{figure}

\Cref{fig:comparison_frankwolfe_momentum} illustrates the impact of using momentum on Frank-Wolfe.
Overall, the relative gap to the best objective is indeed reduced.
As a reference, \cref{fig:comparison_frankwolfe_momentum} also illustrates the gap to \ac{DiffNC} with \ac{SLSQP}, showing also better optimality.

\subsection{Execution time}
\label{sec:eval:execution_time}

\begin{figure}[h!]
	\centering
	\includegraphics[width=\columnwidth]{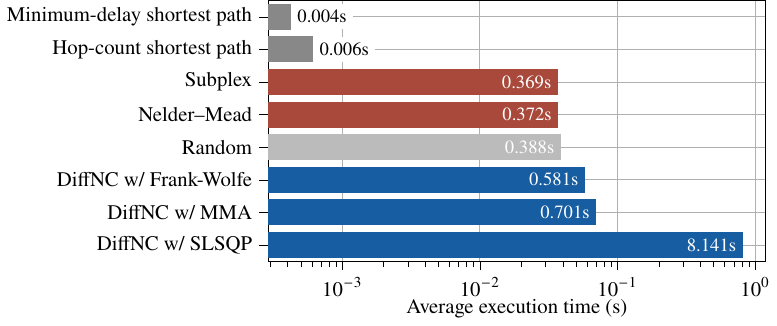}
	\caption{Average execution time to find the optimal solution.}
	\label{fig:ecdf_execution_time}
\end{figure}

Following our discussion on the ways to optimize the computation speed of \ac{DiffNC}, we compare here the execution time of the optimization part of \ac{DiffNC} against the other heuristics.
Results are presented in \cref{fig:ecdf_execution_time}.
Due to the limit of 500 evaluations of the objective function, most of the algorithms exhibit here similar execution times.
The evaluations presented here were performed on an AMD EPYC 7702P with 128 cores with use of the parallelization approach presented in \cref{sec:eval:parallelization}.

\begin{figure}[h!]
	\centering
	\includegraphics[width=\columnwidth]{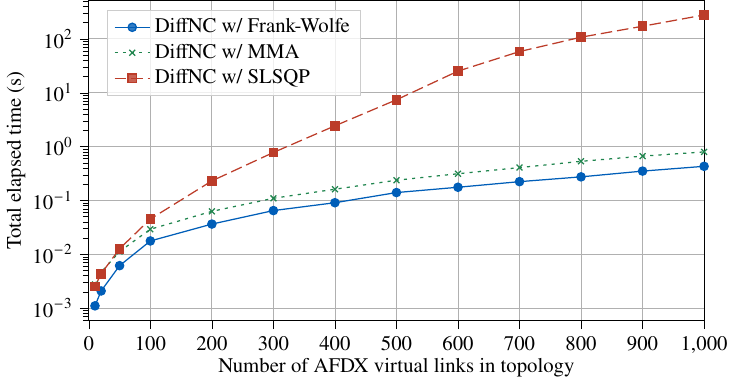}
	\caption{Scalability of \ac{DiffNC} on the \ac{AFDX} network when optimizing for two network-wide priority levels.}
	\label{fig:afdx_execution_time_scalability}
\end{figure}

One outlier is \ac{SLSQP} which is one order of magnitude slower than the other methods.
This is due to the fact that its execution time grows quadratically with the number of variables of the optimization problem.
To illustrate this issue, we used \ac{DiffNC} on the \ac{AFDX} network, where we incrementally increase the number of virtual links in the network.
Results are presented in \cref{fig:afdx_execution_time_scalability}.
\ac{DiffNC} with \ac{SLSQP} is almost 3 orders of magnitude slower than \ac{DiffNC} with Frank-Wolfe on the full network, showing its poor scalability.

\begin{figure}[h!]
	\centering
	\includegraphics[width=\columnwidth]{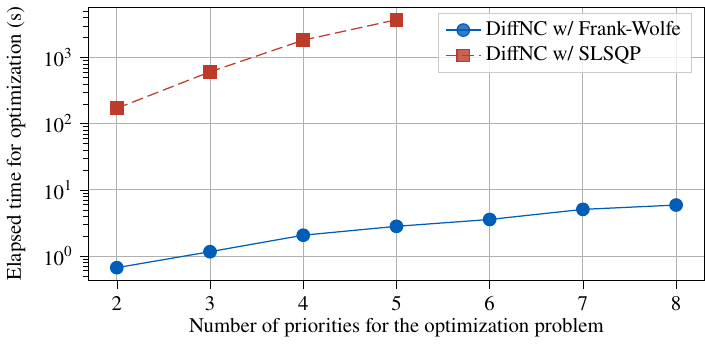}
	\caption{Computation time required for optimizing the \ac{AFDX} network using \ac{DiffNC} w/ Frank-Wolfe and \ac{SLSQP}. Results for \ac{SLSQP} with more than 5 priorities are omitted here since they take more than \SI{10}{\hour} to compute.}
	\label{fig:execution_time_vs_nprios_a350_frankwolfe}
\end{figure}

To further illustrate the scalability of \ac{DiffNC} with Frank-Wolfe, we optimized the \ac{AFDX} network with an increasing number of (network-wide) priorities.
Results are presented in \cref{fig:execution_time_vs_nprios_a350_frankwolfe}.
The execution time grows linearly with the number of priorities.
Compared to \ac{DiffNC} with \ac{SLSQP}, \ac{DiffNC} with Frank-Wolfe is three orders of magnitude faster.

Overall, our evaluations show that \ac{DiffNC} with Frank-Wolfe is an efficient method for optimizing networks under delay bound constraints, outperforming all the other optimization methods evaluated here, and at a reasonable computational cost.
This also applies to large real networks with more than \num{1000} flows, illustrating that this method scales to real industrial networks.

%% file: conclusion.tex
\section{Conclusion}
\label{sec:conclusion}

We introduce in this article \ac{DiffNC}, an extension of \ac{NC}, by showing that the (min,plus)-algebraic terms derived by \ac{NC} are differentiable.
The term bounding the end-to-end delay of a flow can already be differentiated w.r.t. to curve parameters or flow priorities.
However, our approach also allows for network design and synthesis.

We investigate the optimization of flow paths and priority assignment, a task known to be difficult due to its combinatorial nature.
An extension of \ac{NC} models to include alternative flow paths allows to differentiate w.r.t. these.
We show that \ac{DiffNC} with variable relaxation is able to reformulate the optimization problem as a constrained nonlinear optimization problem that can be optimized using gradient-based methods.
Our numerical evaluation shows that \ac{DiffNC} with Frank-Wolfe can reduce the average delay bounds by \SI{39.2}{\percent} compared to shortest path routing.

We also show that our approach is able to scale to industrial networks.
We demonstrate that \ac{DiffNC} with Frank-Wolfe is able to optimize the \acl{AFDX} network from the Airbus A350 in a matter of seconds, outperforming \ac{DiffNC} with \acl{SLSQP} by several orders of magnitude.
Furthermore, a comparison with other optimization methods for combinatorial and nonlinear optimization shows that \ac{DiffNC} can outperform global search methods.

%% file: bibliographies.tex
\begin{IEEEbiographynophoto}{Fabien Geyer}
is currently with Airbus Central Research \& Technologies and Technical University of Munich (TUM) working on methods for network analytics, network performances and architectures. He received the master of engineering in telecommunications from Telecom Bretagne, France in 2011 and the Ph.D. degree in computer science from TUM in 2015. His research interests include novel methods for data-driven networking, formal methods for performance evaluation and modeling of networks.
\end{IEEEbiographynophoto}

\begin{IEEEbiographynophoto}{Steffen Bondorf}
is the Professor of Distributed and Networked Systems in the Faculty of Computer Science
at Ruhr University Bochum, Germany. 
Steffen received his Dr.-Ing. in Computer Science from TU Kaiserslautern, Germany, in 2016.
After graduation, he was a research fellow at National University of Singapore and an ERCIM Fellow at NTNU Trondheim, Norway.
Steffen's research interests are in performance analysis of networked systems.
\end{IEEEbiographynophoto}